\documentclass[useAMS,usenatbib]{mn2e}
\usepackage{graphicx}

\title[Corotation and the high energy synchrotron emission]{Influence of corotation on the high energy synchrotron emission in Crab-like pulsars}

\author[Osmanov Z.]{Osmanov Z.\thanks{E-mail:
z.osmanov@astro-ge.org} \\
Center for Theoretical Astrophysics, ITP, Ilia State University,
Kazbegi ave. 2$^a$, 0160, Tbilisi, Georgia}
\begin{document}

%\date{Accepted 1988 December 15. Received 1988 December 14; in original form 1988 October 11}

\pagerange{\pageref{firstpage}--\pageref{lastpage}} \pubyear{2009}

\maketitle

\label{firstpage}

\begin{abstract}
   For Crab-like pulsars we consider the synchrotron mechanism
   influenced by relativistic effects of rotation to study the production of the
   very high energy (VHE) pulsed radiation.
   The process of quasi-linear diffusion (QLD) is applied to
   prevent the damping of the synchrotron emission due to
   extremely strong magnetic field.
   By examining the kinetic equation governing the QLD, apart from
   the synchrotron radiative force, we taken into account the the so-called
   reaction force, that is responsible for corotation and
   influences plasma processes in the nearby zone of the light cylinder (LC)
   surface. We have found that the relativistic effects of rotation significantly
   change efficiency of the quasi-linear diffusion. In particular,
   examining magnetospheric parameters
   typical for Crab-like pulsars, it has been shown that unlike the situation,
   where relativistic effects of rotation are not important,
   on the LC surface, the relativistic electrons via
   the synchrotron mechanism may produce photons even in the TeV domain.
   It is shown that the VHE radiation is strongly correlated with the relatively
   low frequency emission.
\end{abstract}


\section{Introduction}

According to the standard pulsar model, in particular, the so-called
polar cap model \cite{stur}, particles uproot from the star's
surface, and accelerate along the magnetic field lines
\cite{rudsuth} inside a gap zone. In general, it is believed that
these particles produce the nonthermal radiation, which is
interpreted in terms of the synchrotron mechanism \cite{pacini,shkl}
and the Inverse Compton Scattering \cite{blan} respectively.
Generally speaking, due to strong synchrotron losses, relativistic
electrons quickly lose their perpendicular momentum, and transit to
their ground Landau states. Therefore, particles may be
approximately described as moving one-dimensionally along the field
lines, which in turn means that the synchrotron mechanism does not
contribute in high energy radiation processes. On the other hand,
under certain conditions, due to the quasi-linear diffusion (QLD),
the pitch angles might be recreated, leading to the efficient
synchrotron emission. This method has already been applied to
pulsars \cite{machus1,malmach,nino,nino2,nino1,difus,difus1} and
active galactic nuclei \cite{difus3,difus4}.

In the context of QLD, the recent detection of the VHE radiation
from the Crab pulsar \cite{magic} could be very important. MAGIC
Cherenkov telescope from 2007 October to 2008 February, has
discovered the pulsed emission above 25GeV, which reveals several
characteristic features. A special interest deserves the coincidence
of signals from different energy bands ranging from radio to VHE
($>25$GeV) domain \cite{magic}. The authors conclude that according
to the data, the polar cap models must be excluded from the possible
scenario of the radiation, which must happen far out in the
magnetosphere. Analysis of the MAGIC data also implies that a
location of the aforementioned VHE and low energy emissions must be
the same. By taking into account the synchrotron reaction force, we
constructed the kinetic equation governing the QLD and the results
of the MAGIC data have been interpreted (Machabeli \& Osmanov 2009).
We argued that the observed VHE radiation is produced by the
synchrotron mechanism, having properties, which are in a good
agreement with the observations. The following work (Machabeli \&
Osmanov 2010) was related to the problem of the curvature radiation
and the inverse Compton scattering in the context of the same
observations. We have shown that these mechanisms do not contribute
to the VHE domain detected by MAGIC.

In general, magnetic field in pulsar magnetospheres is huge and
ranges from $10^6$G (far out in the magnetosphere) to $10^{12}$G
close to the neutron star's surface. Therefore, magnetospheric
plasma is in the frozen-in condition and is forced to follow the
magnetic field lines. This means that effects of corotation in
plasma acceleration and emission processes could be very important.
In general, it is believed that the high energy radiation comes from
an area located relatively close to the LC surface, where corotation
is extremely significant. For this purpose it is reasonable to
consider the quasi-linear diffusion by taking relativistic effects
of rotation into account and see how the corotation influences the
mentioned processes.

In the present paper we study the role of the corotation in the
quasi-linear diffusion and we show that in the very vicinity of the
LC surface, contrary to the quasi-linear diffusion, effects of
corotation attempt to decrease the pitch angles, and kill the
subsequent synchrotron emission.

The paper is organized as follows. In Section 2 we consider the
quasi-linear diffusion by taking the effect of corotation into
account, in Sect. 3 we present our results and in Sect. 4 we
summarize them.

%%%%%%%%%%%%%%%%%%%%%%%%%%%%%%%%%%%%%%%%
\section{Main consideration} \label{sec:consid}
%%%%%%%%%%%%%%%%%%%%%%%%%%%%%%%%%%%%%%%%

In this section we present the model, generalizing our previous
approach by taking the fact of corotation in to account and see how
the efficiency of QLD depends on it.

Generally speaking, the pulsar magnetospheres are composed of low
and high energy particles respectively. Therefore, in the framework
of the paper for simplicity we assume that the magnetosphere is
consist of two components: (a) the plasma component with the Lorentz
factor, $\gamma_p$ and (b) the beam component with the Lorentz
factor, $\gamma_b$.

According to the mechanism of the quasi-linear diffusion the
following transverse mode generates
\begin{equation}\label{disp1}
\omega_t \approx kc\left(1-\delta\right),\;\;\;\;\;\delta =
\frac{\omega_p^2}{4\omega_B^2\gamma_p^3},
\end{equation}
where $k$ denotes the modulus of the wave vector, $c$ is the speed
of light, $\omega_p \equiv \sqrt{4\pi n_pe^2/m}$ is the plasma
frequency, $e$ and $m$ are electron's charge and the rest mass,
respectively, $n_p$ is the plasma density, $\omega_B\equiv eB/mc$ is
the cyclotron frequency and $B$ is the magnetic induction. It can be
shown that the aforementioned waves excite if the cyclotron
resonance condition is satisfied \cite{kmm}
\begin{equation}\label{cycl}
\omega - k_{_{\|}}\upsilon-k_xu\pm\frac{\omega_B}{\gamma_b} = 0.
\end{equation}
By $k_{_{\|}}$ and $k_{\perp}$ we denote the wave vector's
longitudinal (parallel to the background magnetic field) and
transverse (perpendicular to the background magnetic field)
components respectively, $u\equiv c\upsilon\gamma_b/\rho\omega_B$ is
the drift velocity of electrons, $\upsilon$ is the component of
velocity along the magnetic field lines and $\rho$ is field line's
curvature radius. In this paper we consider the beam particles as to
be the resonance particles. One can show from Eqs.
(\ref{disp1},\ref{cycl}) that the excited cyclotron wave is
characterized by the following frequency \cite{malmach}
\begin{equation}\label{om1}
\nu\approx \frac{\omega_B}{2\pi\gamma_b\delta}.
\end{equation}

Generally speaking, particles moving in the magnetic field,
experience two forces, one of which, ${\bf G}$, having the following
components \cite{landau}
\begin{equation}\label{g}
G_{\perp} = -\frac{mc^2}{\rho}\gamma_b\psi,\;\;\;\;\;G_{_{\|}} =
\frac{mc^2}{\rho}\gamma_b\psi^2.
\end{equation}
is responsible for the conservation of the adiabatic invariant, $I =
3cp_{\perp}^2/2eB$ and the second one is the synchrotron radiative
force \cite{landau}
\begin{equation}\label{f}
F_{\perp} = -\alpha\psi(1 + \gamma_b^2\psi^2),\;\;\;\;\;F_{_{\|}} =
-\alpha\gamma_b^2\psi^2,
\end{equation}
where $\alpha = 2e^2\omega_B^2/(3c^2)$. But one can show that for
the Crab-like pulsars the radiation reaction force, $|{\bf F}|$,
exceeds $|{\bf G}|$ by many orders of magnitude (Machabeli \&
Osmanov 2009,2010).

On the other hand, apart from the synchrotron radiative force,
relativistic electrons also experience a reaction force responsible
for the corotation\footnote{In the local frame of reference
particles experience the centrifugal force, whereas considering
dynamics in the laboratory frame, one has to examine the reaction
force acting on the electrons from the magnetic field line.}. This
force can be estimated by a simple mechanical analogy introduced by
Machabeli \& Rogava (1994) and reconsidered by \cite{grg}. In the
framework of this approach, instead of magnetic field lines and
particles sliding along them, one considers a corotating pipe with a
bead inside it. If we assume that the magnetic field lines are
straight, then the reaction force acting on a single particle from
the beam component can be given by \cite{grg}
\begin{equation}\label{reac}
R = \frac{dp_{_\perp}}{dt} + \Omega p_{_\parallel},
\end{equation}
where $\Omega$ is the angular velocity of rotation, $r$ is the
radial coordinate,
\begin{equation}\label{pr}
p_{_\parallel}\equiv\gamma_b m\upsilon
\end{equation}
and
\begin{equation}\label{pfi}
p_{_\perp}\equiv\gamma_b mr\Omega
\end{equation}
are momentum's longitudinal and transversal components respectively.

Considering a single particle approach, one can show that the
Lorentz factor of a particle moving along the corotating straight
magnetic field yields the equation \cite{rm00}
\begin{equation}\label{gama}
\gamma_b =  \frac{1}{ \sqrt{\widetilde{m}}
\left(1-\frac{r^2}{r_{lc}^2}\right)},
\end{equation}
where
$$\widetilde{m} =\frac{1-r_0^2/r_{lc}^2-\upsilon_0^2/c^2}{\left(
1-r_0^2/r_{lc}^2\right)^2},\; r_{lc}\equiv \frac{c}{\Omega}.$$
$r_0$ and $\upsilon_0$ are particle's initial position and the
initial radial velocity respectively and $r_{lc}$ is the light
cylinder radius. From Eq. (\ref{gama}) we see that the closer to the
LC, the bigger the Lorentz factor. Therefore a dynamical effect of
the corotation becomes extremely efficient nearby this area [see Eq.
(\ref{reac})] where the reaction force can be approximated as to be
[see Eq. (\ref{r4}) in Appendix]
\begin{equation}\label{reac1}
R\approx 2 m\widetilde{m}^{1/4}c\Omega\gamma_b^{3/2}.
\end{equation}

\begin{figure}
  \resizebox{\hsize}{!}{\includegraphics[angle=0]{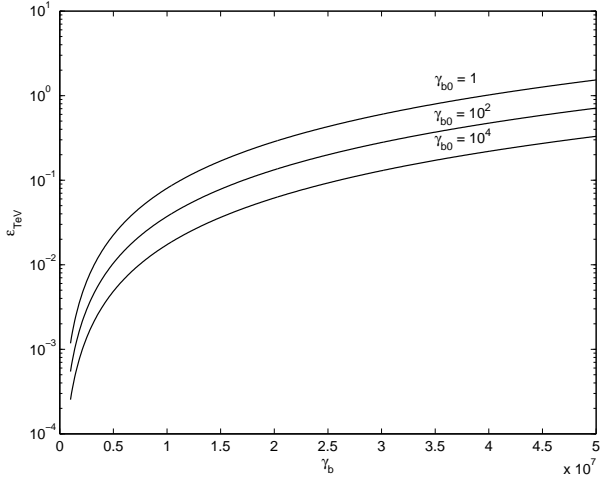}}
  \caption{The synchrotron emission energy versus the Lorentz factors of beam electrons.
  The set of parameters is $P\approx 0.0332$s, $B\approx 1.7\times 10^7$G,
  $\gamma_{b0} = \{1,10^2,10^4$\}. }\label{fig1}
\end{figure}

As we have mentioned, in the Crab pulsar magnetosphere one may
neglect the effects of ${\bf G}$ in comparison with ${\bf F}$.
Therefore, the process of the quasi-linear diffusion is mainly
influenced by the synchrotron radiative force, ${\bf F}$ and the
reaction force, ${\bf R}$. The synchrotron radiation reaction force,
acting on a particle, attempts to decelerate it, therefore
relativistic electrons loose their perpendicular momentum, leading
to an inevitable decrease of the pitch angles. In spite of the fact
that contrary to ${\bf F}$, the reaction force plays an accelerative
role in electron's dynamics, in the context of the pitch angles its
role is the same. In particular, as we have already discussed, the
mentioned reaction force is a direct consequence of corotation. This
means that ${\bf R}$ is constructed in such a way that particles
always must stay on the magnetic field lines, therefore, the
reaction force also attempts to decrease the pitch angles. On the
other hand, the quasi-linear diffusion, that arises through the
influence of the waves back on the particles, via the QLD attempts
to widen a range of the pitch angles. The dynamical process
saturates when the effects of the above mentioned forces are
balanced by the diffusion. For $\gamma_b\psi\gg 1$ it is easy to
show that in the quasi-stationary case ($\partial/\partial t = 0$),
the corresponding kinetic equation governing the QLD writes as
follows \cite{malmach}
\begin{equation}\label{kinet}
\frac{1}{p_{_\parallel}}\frac{\partial\left[\psi\left(R+
F_{\perp}\right)\chi\right]}{\partial\psi}
=\frac{\partial}{\partial\psi}\left[\psi
D_{\perp\perp}\frac{\partial\chi}{\partial\psi}\right],
\end{equation}
where $\chi = \chi (\psi)$ is the distribution function of particles
with respect to the pitch angles,
\begin{equation}\label{dif}
D_{\perp\perp}\approx -\frac{\pi e^2n_bc}{4\nu},
\end{equation}
is the diffusion coefficient, $n_b = B/Pce$ is the density of the
beam component and $P$ is the pulsar's rotation period. The solution
of Eq. (\ref{dif}) writes as follows
\begin{equation}\label{chi} \chi(\psi) = \chi_{_0}e^{-A_1\psi\left(1-A_2\psi^3\right)},\;\;\;
A_1\equiv\frac{mc\gamma_bR}{D_{\perp\perp}},\;\;\;
A_2\equiv\frac{\alpha\gamma_b^2}{4R}.
\end{equation}
%
%\begin{equation}\label{A}

%\end{equation}
%

As is clear from this expression, electrons are differently
distributed for different pitch angles. Let us note that the
reaction force and the synchrotron radiative force have opposite
directions, since $F_{\perp}$ is responsible for deceleration of
electrons, whereas $R$ accelerates them. It is clear that close to
the LC, a value of $|R|\propto \gamma_b^{3/2}$ is very high and for
very small pitch angles exceeds $|F_{\perp}|$ and therefore, higher
values of pitch angles lead to higher values of the distribution
function. But by increasing $\psi$ the corresponding reaction force
does not change [see Eq. (\ref{reac1})], whereas the synchrotron
radiative force is very sensitive to this change,
$F_{\perp}\sim\psi^3$. Therefore, at a certain pitch angle,
$\psi_0$, these forces will balance each other and the distribution
function will reach its maximum value. In particular, one can
straightforwardly show that the expression of the distribution
function [see Eq. (\ref{chi})] peaks at
\begin{equation}\label{pitch}
\psi_0 = \frac{1}{\sqrt[3]{4A_2}} =
\left(\frac{2mc\Omega}{\alpha\gamma_{b0}^2\gamma_b^{1/2}}\right)^{1/3},
\end{equation}
which indeed corresponds to $R=F_{\perp}$. Here, we have taken into
account the following relation $\widetilde{m}\approx
1/\gamma_{b0}^2$ [$r_0/r_{lc}\ll 1$, see Eq. (\ref{gama})].

It is reasonable to estimate how steep is the distribution function.
For simplicity we consider the case $\gamma_{b0} = 1$, then
examining the Crab pulsar's ($P\approx 0.0332$s) magnetospheric
parameters close to the LC surface: $B\approx 1.7\times 10^7$G,
$\gamma_b = 10^7$, one can see that at the peak, $\psi_0\approx
0.02$rad, the distribution function exceeds its value at $\psi = 0$
by many orders of magnitude. This means that most of the particles
will be characterized by the peak value of the pitch angle, and
therefore, its average value can be estimated as to be $\psi_0$. We
have taken into account that $B = B_{st}R_{st}^3/r_{lc}^3$, where
$R_{st}\approx 10^6$cm is the pulsar's radius, $B_{st}\approx
1.8\times 10^{12}\sqrt{P\times\dot{P}_{15}}$G is the magnetic
induction on the star's surface and $\dot{P}_{15}\equiv
10^{15}\dot{P}$ ($\dot{P}_{15}=421$ for the Crab pulsar, see
\cite{manch}).

Since we consider centrifugally accelerated particles, our approach
is based on the assumption that electrons corotate with the spinning
pulsar, which is valid only inside a certain zone (corotation zone
with a radius $r_c$), because in this region the magnetic field is
strong enough to channel the flow. This means that, our approach is
valid if the following condition is satisfied:
$(r_{lc}-r_c)/r_{lc}\ll 1$. It is clear that corotation takes place
if the magnetic energy density exceeds the beam energy density,
therefore, $r_c$ can be can be estimated by the condition:
$B(r)^2/(8\pi)\approx mc^2n_b\gamma(r)$. If we consider a range of
$\gamma_{b0} = (1; 10^4)$, then by taking Eq. (\ref{gama}) into
account, one can see that for $\gamma_b\sim 10^7$ one has
$(r_{lc}-r_c)/r_{lc}\sim (10^{-11}; 10^{-7})\ll 1$, which means that
corotation is violated in a very close region to the LC surface and
for almost the whole course of motion the spinning magnetic field
channels the electrons.

When relativistic particles move in the magnetic field, they emit
electromagnetic waves, which in our case correspond to the following
photon energies \cite{Lightman}:
$$\epsilon_{TeV}\approx 1.7\times
10^{-20}B\gamma^2\sin\psi_0\approx$$
\begin{equation}
\label{eps}  \;\;\;\;\;\;\;\;\;\;\;\;\;\;\;\;\;\;\;\;\;\;\;\;\approx
10^{-16}\times B^{1/3}\times\gamma_b^{11/6}\times\gamma_{0b}^{-2/3}.
\end{equation}

\section{Discussion}

In this section we are going to apply the method to the Crab-like
pulsars and see what changes in the process of the QLD when
corotation is taken into account.

In Fig. \ref{fig1} we show the dependence of photon energies on the
Lorentz factors of the beam electrons for three different initial
values of the beam Lorentz factors. The set of parameters is
$P\approx 0.0332$s, $B\approx 1.7\times 10^7$G, $\gamma_{b0} =
\{1,10^2,10^4$\}. As is clear from the plots by increasing
$\gamma_b$ the corresponding emission energy increases as well. This
is a natural result, because more energetic electrons emit more
energetic photons [see Eq. (\ref{eps})]. A second feature concerns
the behaviour of $\epsilon_{TeV}$ versus $\gamma_{b0}$. It is
evident that the higher the initial beam Lorentz factor the lower
the emission energy. Indeed, as one can see from Eq. (\ref{pitch})
the pitch angle behaves as $\gamma_{b0}^{-2/3}$, therefore, for
higher initial beam Lorentz factors one obtains lower values of the
pitch angles and hence, the corresponding synchrotron energy becomes
lower.
\begin{figure}
  \resizebox{\hsize}{!}{\includegraphics[angle=0]{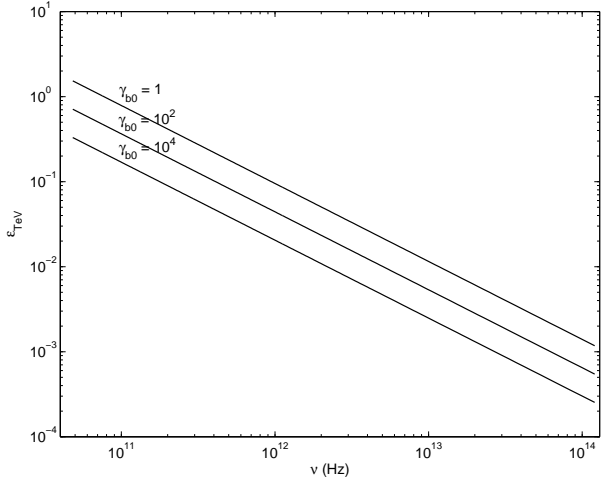}}
  \caption{The synchrotron emission energy versus the cyclotron frequency.
  The set of parameters is $P\approx 0.0332$s, $B\approx 1.7\times 10^7$G,
  $\gamma_{b0} = \{1,10^2,10^4$\}. }\label{fig2}
\end{figure}
The relativistic effects of rotation could also be interesting in
the context of the recently detected VHE pulsed emission ($>25$GeV)
from the Crab pulsar. According to the interpretation presented in
(Machabeli \& Osmanov 2009,2010), the mentioned radiation is formed
on lengthscales of the order of $10^8$cm via the quasi-linear
diffusion. As it has been shown, in order to explain the data, the
value of the beam Lorentz factor must be of the order of $3\times
10^8$. For centrifugally accelerated electrons the effects of
corotation become significant in the nearby zone of the LC surface.
In particular, from Eq. (\ref{reac1}) it is clear that the reaction
force is proportional to $\gamma_b^{3/2}$, which in turn
asymptotically increases by reaching the LC surface [see Eq.
(\ref{gama})]. The corresponding lengthscale of a layer on the light
cylinder zone, where the effects are intensified by corotation, can
be estimated as follows
\begin{equation}
\label{lamb}  \lambda\approx \frac{\gamma_b}{d\gamma_b/dr}\approx
\frac{r_{lc}}{2\gamma_b},
\end{equation}
where we have taken into account Eq. (\ref{gama}). As we see from
this expression, the layer is quite thin, $\lambda\ll r_{lc}$.
Unlike the papers by Machabeli \& Osmanov (2009,2010) where the
effects of corotation are not significant, and the VHE radiation
($>25$GeV) can be achieved by the relativistic particles with
$\gamma_b\sim 3\times 10^8$, a part of the magnetosphere located
very close to the LC surface, may guarantee the aforementioned
emission energy for lower Lorentz factors. In particular, from Fig.
\ref{fig1} it is clear that the synchrotron process supported by the
QLD produces $25$GeV photons for $\gamma_{b}\sim 5.3\times 10^6$
($\gamma_{ b0} = 1$), $\gamma_{b}\sim 8\times 10^6$ ($\gamma_{b0} =
10^2$) and $\gamma_{b}\sim 1.2\times 10^7$ ($\gamma_{b0} = 10^4$),
which are much lower than $3\times 10^8$. The quasi-linear diffusion
is so efficient that under certain conditions it can provide TeV
photons as well. As one can see from the figure, the relativistic
electrons with Lorentz factors $5\times 10^7$ ($\gamma_{b0} = 1$)
may provide TeV emission ($1.5$TeV).

According to the present method, the process of the QLD depends on
excitation of the unstable cyclotron waves, which correlate with the
high energy radiation. In Fig. \ref{fig2} we show the behaviour of
synchrotron emission energy versus the cyclotron frequency. The set
of parameters is the same as in Fig. \ref{fig1}. As is seen from the
plots, the VHE radiation is strongly connected with relatively low
energy radiation starting from microwave ($5\times 10^{10}$Hz) to
optical ($10^{14}$Hz) domains respectively.

The investigation shows that very close to the LC area, where
effects of corotation are extremely important, the quasi-linear
diffusion becomes so efficient that under favorable conditions it
may provide VHE radiation for relatively lower energy electrons than
in the regime considered by Machabeli \& Osmanov (2009,2010). The
aim of the present paper was to demonstrate for Crab-like pulsars
that the relativistic effects of rotation are of fundamental
importance for studying the VHE emission via QLD. Studying
rotationally driven quasi-linear diffusion for $1$-sec pulsars also
is an interesting problem, but it is out of the scope of the present
paper and sooner or later we are going to consider it as well.

\section{Summary}\label{sec:summary}

\begin{enumerate}

      \item We have examined the VHE radiation of Crab-like
      pulsars via the quasi-linear diffusion intensified by the effects
      of corotation which takes place in the nearby zone of the LC surface.

      \item It has been emphasized that due to very strong magnetic
      field of pulsar magnetospheres the efficient energy losses lead
      to the
      damping of the synchrotron process. Close to the LC surface
      the effects of corotation become significant
      and apart from the synchrotron radiative force, which decreases
      the pitch angles, also the reaction force (responsible for
      corotation) has to be taken into account. Generalizing the
      kinetic equation governing the QLD, we have found the particle
      distribution with respect to $\psi$ and estimated its average
      value.

      \item Considering Crab-like pulsar's magnetospheric parameters we have
      found that the quasi-linear diffusion becomes more efficient on
      the LC zone, than for locations far from this area.
      It has been shown that the synchrotron emission process, via the QLD,
      might provide VHE emission even in the TeV domain.

      \item We have found that the VHE radiation is strongly connected with
      the cyclotron emission having relatively low frequencies.

      \end{enumerate}

%\acknowledgments
\section*{Acknowledgments}
The research was supported by the Georgian National Science
Foundation grant GNSF/ST07/4-193.

\appendix

\section{}

In this section we would like to estimate the reaction force close
to the LC surface.

From Eqs. (\ref{reac}-\ref{pfi}) one can see that
\begin{equation}
\label{r1} R = m\Omega\frac{d\left(\gamma_b r\right)}{dt}+m\gamma_b
\upsilon\Omega = m\Omega
r\frac{d\gamma_b}{dt}+2m\gamma_b\upsilon\Omega.
\end{equation}
By applying Eq. (\ref{gama}) for the nearby zone of the LC area,
$r\approx r_{lc}$, the aforementioned expression writes as follows
\begin{equation}
\label{r2} R\approx
2m\widetilde{m}^{1/2}\upsilon\Omega\gamma_b^2+2m\gamma_b\upsilon\Omega,
\end{equation}
which by taking into account a natural relation
$\widetilde{m}^{1/2}\gamma_b\gg 1$ reduces to
\begin{equation}
\label{r3} R\approx 2m\widetilde{m}^{1/2}\upsilon\Omega\gamma_b^2.
\end{equation}
Machabeli \& Rogava (1994) showed (see Eq.10 in the corresponding
article)
\begin{equation}
\label{vr1}
\upsilon=c\sqrt{\left(1-\frac{r^2}{r_{lc}^2}\right)\left[1-
\widetilde{m}\left(1-\frac{r^2}{r_{lc}^2}\right)\right]},
\end{equation}
which, applying to the LC surface, by using Eq. (\ref{gama}),
obtains the following approximate form
\begin{equation}
\label{vr2} \upsilon\approx
c\sqrt{\left(1-\frac{r^2}{r_{lc}^2}\right)} =
\frac{c}{\widetilde{m}^{1/4}\gamma_b^{1/2}},
\end{equation}
and reduces Eq. (\ref{r3}) to the final expression
\begin{equation}
\label{r4} R\approx 2m\widetilde{m}^{1/4}c\Omega\gamma_b^{3/2}.
\end{equation}

\bsp

\label{lastpage}

\end{document}